# An efficient algorithm for generating AoA networks

Nasser Eddine Mouhoub [1], Abdelhamid Benhocine[2]

[1] Computer science department Bordj Bou Arreridj University, Algeria
Tel.: +213 771 64 98 17    Fax: +213 35 66 63 10
mouhoub_n@yahoo.fr

[2] Information System department, Qassim University, Saudi Arabia
Abdelhamid-benhocine@yahoo.fr

**Abstract**
The activities, in project scheduling, can be represented graphically in two different ways, by either assigning the activities to the nodes '*AoN*' directed acyclic graph (dag) or to the arcs '*AoA* dag'.
In this paper, a new algorithm is proposed for generating, for a given project scheduling problem, an Activity-on-Arc dag starting from the Activity-on-Node dag using the concepts of line graphs of graphs.

## 1. Introduction

Scheduling of project has received a great attention by researchers these last decades. Several fundamental techniques are largely applied and supported by many commercial systems in scheduling. Among these techniques we find Gantt diagram, PERT/ CPM etc …
The Project Evaluation and Review Technique (PERT) has been used as a tool for project management for over four decades. It consists of planning, designing, and implementing a set of activities to accomplish a particular goal or task. For many years, two of the most popular approaches to project management have been the Critical Path Method (CPM) and the Project Evaluation and Review Technique (PERT). J.E. Kelly of Remington-Rand and M.B. Walker of Dupont developed the CPM in the 1950's to assist in scheduling maintenance shutdowns of chemical processing plants. PERT was developed shortly after by the U.S. Navy to manage the development of the Polaris missile [1].
While CPM is a technique to manage projects with deterministic times, PERT is a technique used in projects that has activities with stochastic times [2]. However, practitioners now commonly use the two names interchangeably, or combine them into the single acronym PERT/CPM.
The original versions of PERT and CPM had some important differences. However, they also had a great deal in common, and the two techniques have gradually merged further over the years. In fact, today's software packages often include all the important options from both original versions.
Another similar method to the two quoted above, was developed in France by Roy [3], and called the 'Méthode des Potentiels Metra' (MPM), with the main difference that each activity is represented in the network by a node.
The activities of a project are often constrained by conditions such as "activity $v$ cannot start until activity $u$ has finished". Assuming that no activity is repeated, we can define a precedence relation $\prec$ on the activities, so that $u \prec v$ means that $u$ must finish before $v$ starts. According to Micha et al. [4], the relation $\prec$ can be represented graphically in two different ways, by either assigning the activities to the nodes or to (a subset of) the arcs. In either case, a *directed acyclic graph (dag)* is defined. In an *activity on node* (*AoN*) dag or (MPM network), each activity corresponds one-to-one with a node, and we say that $u \prec v$ is *represented* if there is a directed path of arcs leading from $v's$ node to $u's$ node. Thus, an *AoN* dag is unique except for possible transitive arcs. In an *activity on arc* (*AoA*) dag or (PERT network), each activity $v$ corresponds to an arc, where parallel arcs that share the same start and terminal nodes are permitted. We say $u \prec v$ is represented in an *AoA* dag if there is a path from $t_u$ the terminal node of the arc for $u$, to $s_v$, the start node of the arc for $v$ (the path is empty if $t_u = s_v$).
Experts prefer to work with the PERT network rather than the MPM. This is why according to Fink et al. [5], it is more concise. Furthermore, Hendrickson et al. [6] explains that it is close to the famous Gantt diagram. According to Cohen et al. [7], the structure of the PERT network is much more suitable for certain analytical techniques and optimization formulations. However, the major disadvantage of this method is in the existence of dummy arcs (see Fig. 1 and 2). Their number is likely to be significantly high especially if the size of the network is too large. The *AoA* dag is not unique.

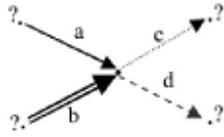

Fig. 1. The representation problem in *AoA* dag.

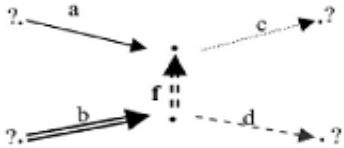

Fig. 2. Introduction of the dummy arc f and the representation in the *AoA* dag

The following example, with only precedence constraints, in Table 1 (the designations are not registered in the schedule table) presents a new technique of generating the *AoA* dag. It is rather a pedagogical method. Generally, the practitioner found difficulties to draw the graph and to mount this difficulty we note that even the experts of project scheduling can not draw correctly. This example which was taken from [5] presented a false *AoA* dag where the schedule table is not respected. The new proposed method uses successive stages in the construction of the *AoA* dag. This is done by scanning the table according to the codes column and the corresponding task to the current scanned line is added taking into account the principle of anteriority (Fig.3.). All the vertices are organized in levels.

We will see later another construction technique of PERT/CPM graph with a bipartite by using the results on the line graphs.

Table 1 : Schedule table

| Codes | Durations | Predecessors |
|---|---|---|
| α | 0 | - |
| A | 2 | α |
| B | 2 | α |
| C | 2 | H |
| D | 3 | α |
| E | 4 | B,G |
| F | 2 | C,I |
| G | 3 | A,D |
| H | 4 | B,D |
| I | 5 | H |
| J | 3 | C |
| ω | 0 | E, J, F |

Table 2: The vertices organized in levels

| Level | Vertices |
|---|---|
| level I | α |
| level II | A , B, D |
| level III | G, H |
| level IV | C, E, I |
| level V | F, J |
| level VI | ω |

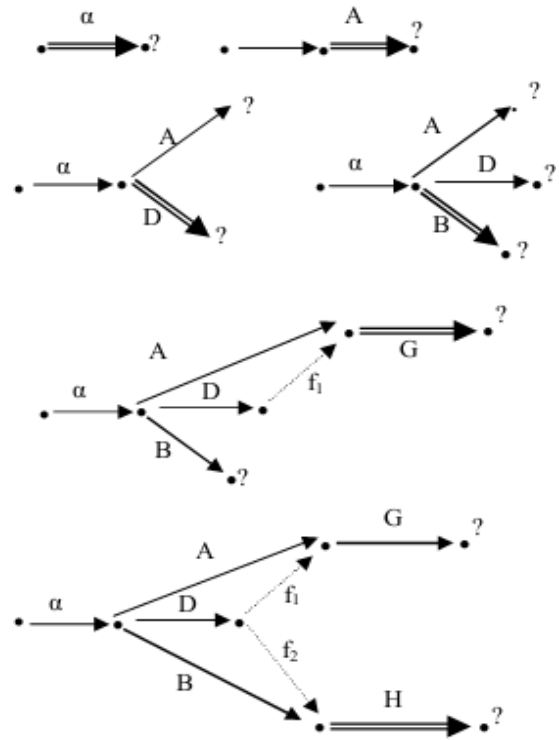

Fig. 3 (a) Constructing *AoA* dag.

and so on until the last graph :

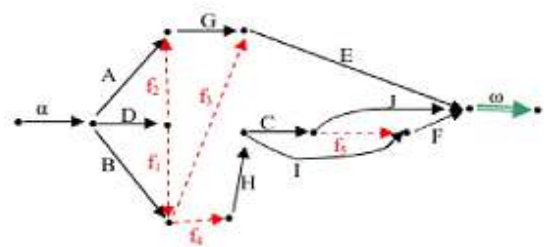

Fig. 3. (b) The *AoA* dag of the Table 1.

We can remind that the durations are not mentioned on the different graphs. These durations can be uncertain. For this precise case, there are more details in [8], [9], [10].
By using the concepts of line graph of graph, we build an *AoA* dag starting from an *AoN* dag. This idea is transformed into an algorithm using graph theory.

## 2. Line graph

Let G = (X, U) a simple or multiple digraph. We build starting from G a graph or line graph noted L (G), called line graph or line digraph of G as follows:
- The nodes of L (G) are in bijective mapping with the nodes of G for simplicity reasons; we give the same name to the arcs of G and the corresponding nodes of L (G).
- Two nodes $u$ and $v$ of L (G) are connected by an arc of $u$ towards $v$ if and only if the arcs $u$ and $v$ of G are such as the final end of $u$ matches with the initial end of $v$, i.e. $T(u) = I(v)$ [11].

### 2.1 Example

Let G the following directed acyclic graph be (Fig 4.):
By definition, any directed graph G admits a unique line graph L (G). On the other hand, two non isomorphs directed acyclic graphs can have the same line graph.

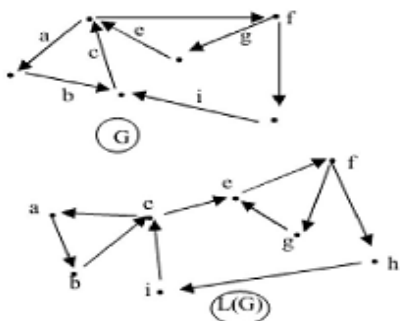

Fig. 4. A graph G and his line graph L(G).

### 2.2 The opposite problem

We suppose the following opposite problem:
Being given a directed acyclic graph H, is it the line graph of any directed acyclic graph? In other words, does there exist a graph G such as L (G) is isomorphs with H, where H = L (G)?

### 2.3 Configuration "Z "and "Δ"

G admits a configuration "Z" if G contains four nodes *a, b, c* and *d* such as if *(a, c), (b, c)* and *(b, d)* are arcs of G, then *(a, d)* is not an arc of G. With an only aim of simplicity, one will give the name of bar of "Z" the arc *(b, c)*.
Configuration "Z" appears when two nodes have common successors and no common successors or by symmetry when two nodes have common predecessors and no common predecessors.

G admits a configuration "Δ" if G contains arcs *(a, b), (b, c)* and *(a, c)* (Fig. 5.) [11].

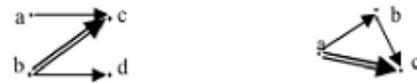

Fig. 5. The configuration "Z" and "Δ".

### 2.4 Some characterizations of the line graphs

The line graphs have been studied but we will present, in this section, the features in which we are interested and obtained from [12]

**1.** H is the line graph of a directed acyclic graph if and only if H does not contain any "Z" configuration.
**2.** H is the line graph of a directed acyclic graph G if and only if arcs of H can be partitioned in a complete bipartite $B_i = (X_i, Y_i)$, i=1..., m, such as $X_i \cap X_j = \phi$ and $Y_i \cap Y_j = \phi$ $\forall$ $i \neq j$.
The bipartite $B_i$ of H are then in a bijection with the nodes also noted $B_i$ which are neither sources nor well, two nodes $B_i$ and $B_j$ of G being connected by an arc from $B_i$ towards $B_j$ if and only if the complete bipartite $B_i$ and $B_j$ of H are such as $Y_i \cap X_j = \phi$ (Fig. 6).
**3.** H is the line graph of a directed acyclic graph without loops if and only if H does not contain any configuration Z or Δ.
**4.** H is the line graph of a directed acyclic graph if and only if any pair of nodes having common successors has all their common successors.
**5.** H is the line graph of a directed acyclic graph if and only if any pair of nodes having common predecessors has all their common predecessors.

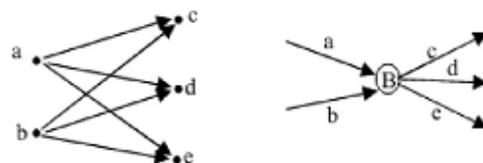

Fig. 6. A complete bipartite B of G and the star of G.

Thus, H is not the line graph of any directed acyclic graph if is only if there is a pair of nodes having common successors and no common successors or common predecessors and no common predecessors [12].

So, we want to know how to transform H in order to get a new graph which is a line graph of a graph.

## 3. Passage from an *AoN* dag to an *AoA* dag

Because of the facility of the use of *AoA* dag, we must concentrate our efforts on the study of the possibility of transforming the *AoN* dag (a significant number of arcs) to *AoA* dag (a reduced number of arcs).
So, we want to know how to transform the graph H (which is an *AoN* dag) in order to get a new graph which is the line graph (*AoA* dag). The difficulty which arises is to know if H does contain Z configurations or not? If it does not contain Z, it is a line graph and the transformation is immediate. But if it contains Z, we have to eliminate the bar from each Z preserving the constraints of succession. Let us study each case apart:

### 3.1 The *AoN* dag is a line graph

Let us build the *AoA* dag starting from the *AoN* dag, assuming that it is a line graph. Under the terms of the results of paragraph 2.4, we proceed as follows:
We partition arcs of the *AoN* dag in a complete bipartite $B_i = (X_i, Y_i)$.

In the *AoA* dag which we wants to build, each $B_i$ is represented by a node still noted $B_i$ and will be the centre of the star.

### 3.2 The *AoN* dag is not a line graph

The construction of *AoA* dag is however more complex in general where the *AoN* dag is not a line graph: it does not admit a partition of the complete arcs in bipartite. It is in this case that one must modify it in order to transform it into a directed acyclic graph associated by preserving the constraints of anteriorities.

Let us suppose that the activities $a_1, ... , a_m$ precede the activities $b_1,... , b_n$. In the *AoN* dag, these constraints of anteriority are represented by a complete bipartite. In *AoA* dag, they are represented by a star (see Fig. 7).

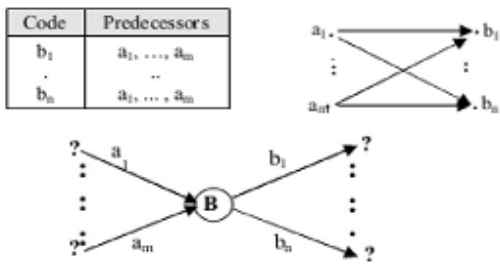

Fig. 7. The sub-table of anteriorities, the complete bipartite B in the line graph and the node B of the corresponding *AoA* dag.

Let us return to the problem of dummy arc in *AoA* dag. If there are for example 4 activities have *a, b, c* and *d* with the following constraints of anteriority: c is preceded by *a* and *b*, but *d* is preceded by *b* only.

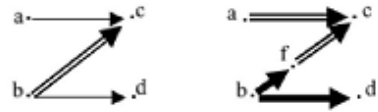

Fig. 8. "Z" configuration, his transformation in *AoN* dag and the partition of completes bipartite.

In *AoN* dag, there is no problem for representing these activities. We do it as in figure 8. But, if there is sub-graphs Z in *AoN* dag (which is considered like the line graph *H*), we are obliged to eliminate all "Z" configuration. We then introduce, in the *AoN* dag, a dummy arc *f* in every "Z" (Fig. 8).

The introduction of the dummy arcs aims at eliminating all "Z" configurations from the *AoN* dag, the constraints remain unchanged. We should recall that the dummy arcs are not necessary in the *AoN* dag but are introduced only to build *AoA* dag.

Kelley [13] notes that it is advantageous to reduce the length of calculations to build an *AoA* dag having the minimum number of nodes and dummy activities.
We then pose the problem of looking for "Z" in the *AoN* dag, i.e. nodes having common successors and no common successors or nodes having common predecessors and no common predecessors.

A first technique of elimination of Z in $G_v$ consists of replacing the bare *(b, c)* of every "Z" by two arcs *(b, f)* and *(f, c)*, according to fig. 8, it is the simplest but the worst, the number of "Z" can be arbitrarily large.

Another technique more effective than the preceding one eliminates several "Z" at the same time by regrouping bars having the initial end or the final end in the same bipartite complete (Fig. 9 and 10).

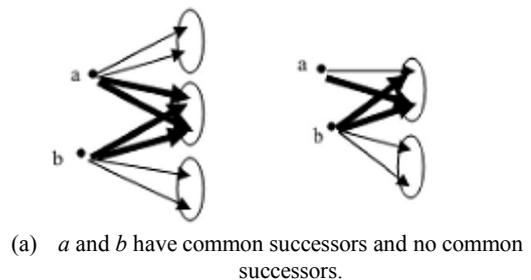

(a)  *a* and *b* have common successors and no common successors.

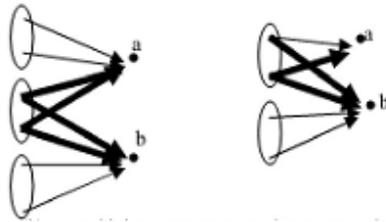

(a) *a* and *b* have common predecessors and no common predecessors.

Fig. 9. Grouping of the "Z" bars having the same initial extremity and belonging to the same complete bipartie

"Z" are destroyed as follows (see Fig. 10):

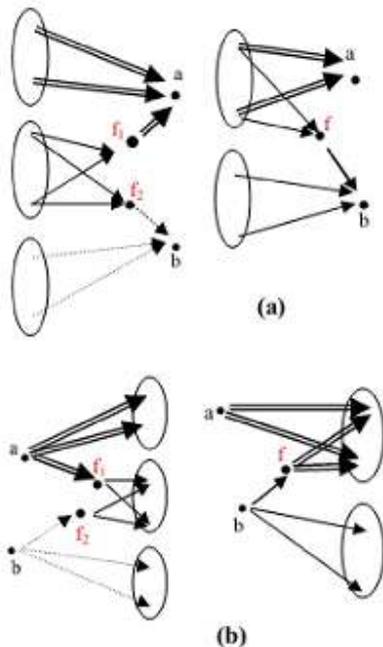

Fig. 10. Grouping of "Z" bars, dummy activities and the arcs partitioning in a complete bipartite.

The problem raised by Kelley [13] was approached by some other authors; more details in [14], [15], [7].

## 4. Algorithm

Let $G_v$ be an *AoN* dag which must be directed, valued, connected and without circuit. With $G_v$ being a conjunctive graph, organized in levels. We want to build the corresponding *AoA* dag which is called $G_e$.

***Begin***
   ***if*** $G_v$ contains Z configurations ***then***
      Identify the $Z_i$ *(i = 1, 2, ..., m)*
      ***For*** every bipartite containing 2 and more bars of Z ***do***
- Create dummy nodes corresponding to the number of bipartite.
- Regroup the bars having same initial extremity or same terminal extremity to the same complete bipartite.
   ***Endfor***
***Endif***
- Identify again the bipartite in $G_v$
- Represent every bipartite $B_i$ in $G_v$ by a node $B_i$ in $G_e$.
- Represent every arc such as :
An arc is drawn between 2 nodes $B_i$ and $B_j$ in $G_e$ if and only if the 2 bipartite $B_i$ and $B_j$ in $G_v$ are such as $Y_i \cap Y_j = \phi$ .

***End.***

### 4.1. Example

Let us consider the following constraints table (Table 3) [16] according to the associated *AoN* dag (Fig. 11. (a)). The designations and the durations are not registered in the schedule table.

Table 3: Schedule table

| Code | Predecessors |
|---|---|
| α | - |
| A | α |
| B | α |
| C | A, B |
| D | A, B |
| E | B |
| F | D |
| G | D |
| H | D, E |
| I | C, F |
| J | C, F, G |
| K | G, H |
| L | J, K |
| ω | I, L |

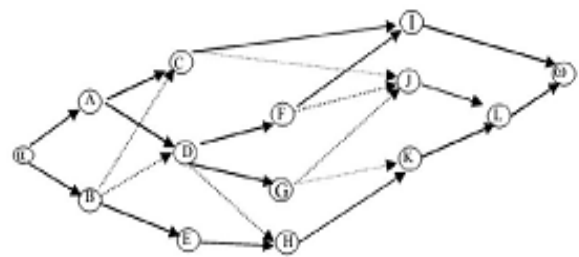

Fig. 11. (a) The *AoN* dag with "Z" bars in dots.

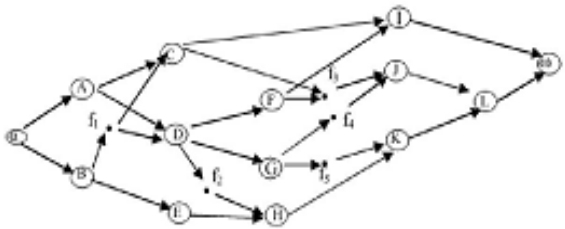

Fig. 11. (b) The modified *AoN* dag by introduction of dummy arcs $f_i$.

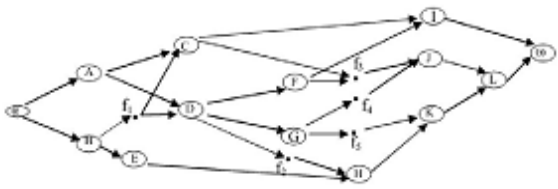

Fig. 11. (c) The *AoN* dag modified with reorganization of the activities in levels and the arcs partitioning in a complete bipartite.

$B_1 = (\{\alpha\}, \{A, B\})$,  $B_2 = (\{B\}, \{F, E\})$,
$B_3 = (\{A, f_1\}, \{C, D\})$,  $B_4 = (\{(C, F\}, \{F, G, f_2\})$,
$B_5 = (\{E, f_2\}, \{H\})$,  $B_6 = (\{C, F\}, \{I, f_3\})$,
$B_7 = (\{G\}, \{f_4, f_5\})$,  $B_8 = (\{f_3, f_4\}, \{J\})$,
$B_9 = (\{f_5, H\}, \{K\})$,  $B_{10} = (\{J, K\}, \{L\})$,
$B_{11} = (\{I, L\}, \{\omega\})$.

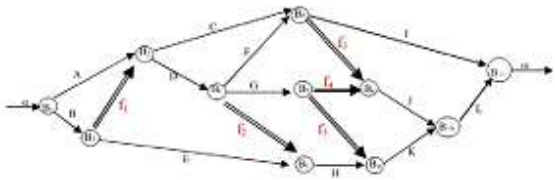

Fig. 11. (d). Construction of an *AoA* dag starting from an *AoN* dag of fig. 11. (a).

## 4.2. Discussion

The algorithm finishes since the loop is carried out only when there is Z, and the number of Z in $G_v$ is known and finite. The rest of the algorithm is a succession of simple instructions. The complexity of the algorithm is polynomial ($O(n^4)$).

## 5. Conclusion

This paper presents a new approach for generating PERT networks. Indeed, by applying some results of line graphs of graphs, the algorithm generates a PERT network with the total respect of the constraints in schedule table.

Our new approach is very simple to be applied. It was programmed in C++ Builder and gives an *AoA* dag in a very short time since its complexity is polynomial $O(n^4)$. The techniques used in the algorithm can be exploited in other fields by specialists in the graph theory. The experimental results are very positive even when networks are of a very large size. Another major benefit that is worth noting is the fact that the algorithm works without any problem in the presence of transitive arcs in *AoN* dag. The new algorithm, due to its new characteristics, is more practical and more user friendly to use by project managers and other practitioners.

This work gives the possibility of new perspectives, such as the treatment of the temporal constraints as well as the research of minimal *AoA* dag in terms of nodes and dummy arcs number.